\documentclass{ws-p10x7-mod2}
\def \as {\alpha_s}

\begin{document}

\title{Power Corrections in Eikonal Cross Sections}

\author{Eric Laenen}

\address{NIKHEF Theory Group, Kruislaan 409\\ 1098 SJ Amsterdam, The
Netherlands}

\author{George Sterman}

\address{C.N.\ Yang Institute for Theoretical Physics,
SUNY Stony Brook\\
Stony Brook, New York 11794 -- 3840, U.S.A.\\
\ \\
Physics Department, Brookhaven National Laboratory,\\
Upton, NY 11973, USA
}

\author{Werner Vogelsang}

\address{RIKEN-BNL Research Center, Brookhaven National Laboratory,\\
Upton, NY 11973, U.S.A.}

\twocolumn[\maketitle\abstract{
We discuss power corrections 
associated with the infrared behavior of the perturbative running
coupling in the 
eikonal approximation to 
Drell-Yan and other annihilation cross
sections in hadron-hadron scattering.  General
properties of the eikonal approximation
imply that  only even powers
of the energy scale are necessary.}]

\section{Introduction}

Power corrections \cite{pc,yd}
are phenomenologically significant in many  QCD hard-scattering
cross sections for which the operator product expansion
is not directly available.  Examples that have received
considerable attention include event shapes in electron-positron
annihilation and transverse momentum distributions in
Drell-Yan cross sections.   In each of these cases, a
perturbative description of the cross section
leads to integrals of the form 
$I_p\equiv Q^{-p}\int_0^Q d\mu\, \mu^{p-1}\, \alpha_s(\mu)$
with $Q$ the hard scale and $p\ge 1$.  In perturbation theory with a fixed
coupling, $I_p$ is just a number, but when the coupling
runs, the integral becomes ill-defined at its lower limit.
This observation requires us to introduce a minimal set
of power corrections 
of the form $\lambda_p/Q^p$, one for
each ambiguous $I_p$ that we encounter \cite{const,irr}.  
The perturbative
expression is cut off, or otherwise regularized to make it
finite without changing the set of exponents $p$.  The values
of the coefficients $\lambda_p$ are then to be found by
comparison with experiment; they depend on the
nature of the perturbative regularization that is employed.
In any case, it is only the sum of regularized perturbation
theory and power corrections that has physical meaning.

The first step in this process is to show that in some
self-consistent approximation the cross section at hand
may be written in terms of integrals like the $I_p$ above.
In many cases, this step involves the resummation of 
logarithms associated with soft gluon emission, for which
the eikonal approximation is useful.  
In this talk\footnote{Contributed to the 30th International Conference on 
High-Energy Physics (ICHEP 2000), Osaka, Japan, 27 Jul - 2 Aug 2000}, 
we discuss an expression for the eikonal approximation
in hadronic collisions, where the analysis of power corrections
through the running coupling is particularly transparent.

\section{The Eikonal Cross Section}

To be specific, we discuss the eikonal approximation
as it appears when
partons $a$ and $b$ combine through an electroweak current,
such as the Drell-Yan annihilation of quark with antiquark
to a lepton pair
or gluon fusion to a Higgs boson,
\begin{equation}
\sigma^{\rm (eik)}_{ab}(q)= \int d^4x\ {\rm e}^{iq\cdot x}
\langle 0 | W_{ab}^\dagger(x) W_{ab}(-0) |0\rangle.
\label{eiksig}
\end{equation}
The operators $W_{ab}$ are defined by
\begin{equation}
W_{ab}(0) \equiv \Phi_{\beta'}^\dagger\, (0)\Phi_\beta(0)\, ,
\end{equation}
in terms of nonabelian phase operators for
$a$ and $b$,
 $\Phi_\beta(0)=P\;{\rm e}^{-ig\int_0^\infty d\lambda \beta\cdot 
A(\lambda\beta)}$, with lightlike velocities $\beta$ and $\beta'$,
$\beta\cdot\beta'=1$.

The eikonal cross sections 
reproduce the logarithms, as
singular as $(Q/q_0)\alpha_s^n$ $\ln^{2n-1}(q_0/Q)$ and
$(Q/q_T)\alpha_s^n\, \ln^{2n-1}(q_T/Q)$, 
that characterize the
edges of partonic phase space at which the
energy of radiation, $q_0$, or its total transverse
momentum, $q_T$, vanish.  The resummation
of these logarithms 
is most convenient in
terms of  transforms, 
\begin{eqnarray}
\tilde \sigma_{ab}^{\rm (eik)}(N,{\bf b})
&=&
\int d^4q\ {\rm e}^{-Nq_0-i{\bf b}\cdot{\bf q}}
\,
\sigma_{ab}^{\rm (eik)}(q)\, .
\nonumber\\
\label{xx}
\end{eqnarray}
In the transformed functions we find
logarithms at each order up to $\alpha_s^n \ln^{2n}N$
and $\alpha_s^n\ln^{2n}(bQ)$, which exponentiate.
The exponentiation of energy logarithms
is known as threshold resummation \cite{dyresum,nkrv,lh}, of transverse
momentum logarithms as $k_T$ resummation \cite{lh,dyqtfact}.

\section{Exponentiation}

Transforms of the eikonal cross section may be
written in exponential form on the basis of 
algebraic considerations that have been known
for a long time,
\begin{equation}
\tilde\sigma_{ab}^{\rm (eik)}(N,{\bf b})
=\exp\; \left[E_{ab}^{\rm (eik)}(N,{\bf b}Q,\epsilon)\right]\, ,
\end{equation}
where the exponent is an integral over functions
$w_{ab}$, sometimes called ``webs" \cite{gath}, which are defined by a 
modified set of
diagrammatic rules,
\begin{eqnarray}
E_{ab}^{\rm (eik)}
&=& 2\, \int^Q {d^{4-2\epsilon}k\over
\Omega_{1-2\epsilon}}\;
\nonumber\\
&\ & \times\
w_{ab} \left(k^2,{k\cdot \beta k\cdot \beta'\over
\beta\cdot\beta'},\mu^2,\alpha_s(\mu^2),\epsilon\right)
\nonumber\\
&\ & \times
\left( {\rm e}^{-N(k_0/Q)+i{\bf k}\cdot {\bf b}}\;-1\right)\, .
\end{eqnarray}
The variable $k$ in this expression may be thought
of as the momentum contributed by the web to the
final state.  The webs factor from each other
under the transforms, and indeed in any symmetric
integral over phase space. \cite{cttw}

Webs have a number of restrictive properties.
At fixed $k$, they are invariant under rescalings of 
the velocities in the eikonal phases, which corresponds to
boost invariance under the axis defined by the two.
In addition, at any fixed order,
the web function has only one overall collinear and IR 
divergence,  from $k_T\rightarrow 0$ and $k_0\rightarrow 0$,
respectively.
Finally, the web functions have no overall renormalization:
\begin{equation}
\mu{d\over d\mu}\; w_{ab} \left(k^2,{k\cdot \beta k\cdot \beta'\over
\beta\cdot\beta'},\mu^2,\alpha_s(\mu^2),\epsilon\right)
=0\, .
\end{equation}

Using  boost invariance in the large-$N$ limit,
we find that the exponent takes the form
\begin{eqnarray}
E_{ab}^{\rm (eik)}
&=&
2\, \int {d^{2-2\epsilon}k_T\over \Omega_{1-2\epsilon}}
\label{EK0}
\\
&\ & \hspace{-7mm} \times
\int_0^{Q^2-k_T^2} dk^2\;
w_{ab} \left(k^2,k_T^2+k^2\right)
\nonumber\\
&\ & \hspace{-7mm} \times
\Bigg[\, {\rm e}^{-i{\bf b}\cdot {\bf k}_T}\;
K_0\left(2N\sqrt{k_T^2+k^2\over Q^2}\right)
\nonumber\\
&\ & \hspace{-3mm} - \ln \sqrt{Q^2\over
k_T^2+k^2}\, \Bigg]
+{\cal O}\left({\rm e}^{-N}\right)\, .
\nonumber
\end{eqnarray}
This expression, which requires dimensional regularization
for its collinear divergences, is completely general 
for the eikonal cross section.

\section{Factorization}

The factorized 
hard-scattering function,
$\hat\sigma_{ab}^{\rm (eik)}$, for the eikonal
cross section
may be constructed in moment space to ${\cal O}(1/N)$ by
dividing by moments of eikonal distributions, \cite{gk}
\begin{eqnarray}
\tilde \phi_{f}^{\rm (eik)}(N,\mu,\epsilon)
&\ & \\
&\ & \hspace{-23mm} =
\exp\left[ - \ln \left( N{\rm e}^{\gamma_E}\right) 
\int_0^{\mu^2}{d\mu'{}^2\over 
\mu'{}^2}\;
A_f \left(\alpha_s(\mu'{}^2)\right)  \right],
\nonumber
\end{eqnarray}
where $A_a$, with
$A_a^{(1)}= C_a$, is the coefficient  of $\ln N$ 
in the $N$th moment of the $a\rightarrow a$ splitting function.
Factorization theorems ensure the cancellation of collinear
divergences in the resulting hard-scattering functions.
Invoking this requirement, we find an explicit
relation  between the webs and the anomalous dimensions,
\begin{eqnarray}
\int_0^{Q^2-k_T^2} dk^2\;
w_{ab} \left(k^2,k_T^2+k^2\right)
&\ &
\\
&\ & \hspace{-40mm}
     =
{A_a\left(\as(k_T^2)\right)+A_b\left(\as(k_T^2)\right) \over
(k_T^2)^{1-2\epsilon}}
+ \dots\, .
\nonumber
\end{eqnarray}

In this fashion, we derive a general form
for the eikonal approximation to the hard-scattering
functions $\hat\sigma_{ab}$ of electroweak annihilation,
\begin{eqnarray}
\hat\sigma_{ab}^{\rm (eik)}(N,{\bf b},Q,\mu)
&=&
{\sigma_{ab}^{\rm (eik)}(N,{\bf b},Q) \over \tilde \phi_a(N,\mu)
\, \phi_b(N,\mu)}
\nonumber\\
&\ &\hspace{-20mm} =
\exp\left[ \hat E_{ab}^{\rm (eik)}(N,b,Q)\, \right]\, ,
\end{eqnarray}
where the collinear-finite exponent (here shown 
in a simplified form, accurate to NLL) is:
\begin{eqnarray}
\hat E_{ab}^{\rm (eik)}
&=&
\int_0^{Q^2} {d k_T^2\over k_T^2}\; \sum_{i=a,b}
A_i\left(\as(k_T^2)\right)\;  \times
\nonumber\\
&\ & \hspace{-15mm}\left[ J_0\left(bk_T\right)
\; K_0\left({2Nk_T\over Q} \right) +
\ln\left({N{\rm e}^{\gamma_E} k_T\over Q}\right) \right]\, .
\nonumber\\
\label{hatE}
\end{eqnarray}
This result is the
basis of the joint threshold-$k_T$ resummation \cite{Liunified} developed
in Ref.\ \cite{lsv}.

As described above, the resummation of logarithms as in (\ref{hatE}) 
requires the inclusion of 
power corrections in both $Q^{-1}$ as well
as $b$, to compensate for the ill-defined behavior of
the strong coupling at low scales.  
In \cite{const} it was shown that 
for the Drell-Yan cross section only integer
powers of $Q^{-1}$ are necessary; in \cite{bbdisp}
models of the running coupling were invoked to suggest that 
power corrections begin at order $Q^{-2}$.  Eq.\
 (\ref{EK0}) implies that only even powers of $Q$ are
present
in all generality for the eikonal approximation.
  This is because, up to a single log, the expansion of the Bessel function 
$K_0(z)$ at small $z$ involves only even
powers of $z$.  
This conclusion includes, and requires, an expression
which, like Eq.\ (\ref{EK0}), is accurate to
the level of ``constant terms'', $(\ln N)^0$.
Other consequences of this approach have
been discussed in Ref.\ \cite{lsv}.

\subsection*{Acknowledgements} 

The work of E.L.\ is part of the research program of the
Foundation for Fundamental Research of Matter (FOM) and the National
Organization for Scientific Research (NWO). 
The work of G.S. was supported in part by the National Science Foundation,
grant PHY9722101. 
G.S.\ thanks Brookhaven National Laboratory for its hospitality.
W.V.\ is grateful to RIKEN, 
Brookhaven National Laboratory and the U.S.  Department of Energy
(contract number DE-AC02-98CH10886) for providing the facilities essential
for the completion of this work.

\end{document}